# LIGHT FIELD RETARGETING FOR MULTI-PANEL DISPLAYS

*Basel Salahieh,\* Seth Hunter, Yi Wu, Oscar Nestares*

Intel Incorporation, Intel Labs, 2200 Mission College Blvd, Santa Clara, CA, USA, 95054
\*E-mail: basel.salahieh@intel.com

## ABSTRACT

*Light fields preserve angular information which can be retargeted to multi-panel depth displays. Due to limited aperture size and constrained spatial-angular sampling of many light field capture systems, the displayed light fields provide only a narrow viewing zone in which parallax views can be supported. In addition, multi-panel displays typically have a reduced number of panels being able to coarsely sample depth content resulting in a layered appearance of light fields. We propose a light field retargeting technique for multi-panel displays that enhances the perceived parallax and achieves seamless transition over different depths and viewing angles. This is accomplished by slicing the captured light fields according to their depth content, boosting the parallax, and blending the results across the panels. Displayed views are synthesized and aligned dynamically according to the position of the viewer. The proposed technique is outlined, simulated and verified experimentally on a three-panel aerial display.*

## 1. INTRODUCTION

Light fields (LFs) [1, 2] are a collection of rays emanating from a real-world scene at various directions, that when properly captured provides a means of calculating depth and parallax cues on 3D displays. A key differentiation between a LF camera and a conventional one is the ability to retrieve angular information (i.e. rays directions) in addition to the spatial content. One way to capture LFs is with plenoptic cameras [3, 4] in which a micro-lens array is added in front of the sensor to preserve the directional component of rays. However, the angular information captured is limited by the aperture extent of the main lens, light loss at the edges of the micro-lenses, and a trade-off between spatial and angular resolution inherent in the design of plenoptic cameras. The resulting multi-view images have a limited baseline, which cannot support a sufficient range of parallax views nor render adequate depth content from different points in the field of view (FOV) of a 3D display.

The data from LF captures can be time-multiplexed and projected onto auto-stereoscopic displays in an additive or multiplicative form. The additive multi-panel displays may consist of multiple layers of rear projection shutters synchronized with projection output at a frequency sufficient to support persistence of vision. Commercial multi-panel displays have been implemented with up to 20 panels [5] spaced 5mm apart in order to support the illusion of a 3D rendered scene inside the display. It is possible to reduce the number of panels required by blending views into a fused representation across the panels [6]. Various LF views can be realized properly on the display by tracking the viewer's position. We consider the 3D display supporting 2D angular view information as LF Display. It is also possible to re-image the light projected on the panels out in front of the display using micro-mirror arrays [7]. The overall effect of adding face tracking and re-imaging is a LF which appears to be continuous and floating in the air relative to the position of the viewer. The multiplicative multi-panel displays on other hand consist of multiple simultaneously active layers that successively modulate the propagated light rays to synthesize the proper views. This has been demonstrated using dual-layer compressive LF display combined with face tracking [8]. Note that the quality of the multiplicative display is highly dependent on the resolution and the refresh rate of the spatial light modulator panels which is not the case for the additive ones since the panels are merely used as rear projection shutters.

Motivated by the idea of a display that could create an aerial LF image with depth information and correct motion parallax cues, we built an additive time-multiplexed three-panel system with re-imaging glass. When the hardware was completed, we displayed static images and solicited initial feedback from viewers in our lab on parallax, planar fusion and the perception of depth in the image. We then used an iterative design process to identify key parameters that would make the floating image more compelling and interactive. In our initial prototype, we were challenged by limited parallax, unmatched resolutions between capture and display, and difficulties mapping 4D LF views to a view-limited 3D display.

In this paper, we present a generalized retargeting pipeline designed to overcome these key perceptual limitations by combining and optimizing the following techniques: 1) a method of synthesizing an enhanced parallax LF content based on estimated depth maps, 2) an algorithm for holes filling using fine slicing, integer shifts, and interpolation, 3) a weighted blending algorithm to achieve continuous depth perception across the panels, and 4) on-the-fly view synthesis enabled by face tracking, angular interpolation, and multi-panel calibration.

The pipeline has been evaluated in simulations and verified experimentally on a 3-panel display. We present the overall rendering pipeline with details on each part of the process and the results on our display. Our contribution is the systematic combination and improvement of the retargeting techniques and the design choices made along the way to achieve efficient light-weight retargeting. We present the pipeline here for other researchers who are retargeting LF data to dynamic multi-panel systems.

## 2. BACKGROUND

Traditionally in LF imaging systems, the displaying device has similar architecture to the capture device to simplify reverse engineering and output LFs naturally representing the captured scenes. An example of similar paired architectures can be found in plenoptic cameras and integral displays. They both utilize a lenslet array to either modulate angular content on the spatial sensor at capture or steer them back to various viewing angles at display. In this paper, we are retargeting LFs that are captured and displayed on completely different architectural devices (see Fig. 1). The capture device in this instance is a plenoptic camera recording 4D coarsely sampled LFs (with angular and spatial content) while the displaying multi-panel device is a 3D display (with coarse depth representations). Among the challenges in such retargeting: 1) the mismatch in resolution (e.g. spatial and angular content) and dimensionality (e.g. mapping 4D content onto 3D display), and 2) the need to enhance the parallax content of captured LFs for a better displaying experience.

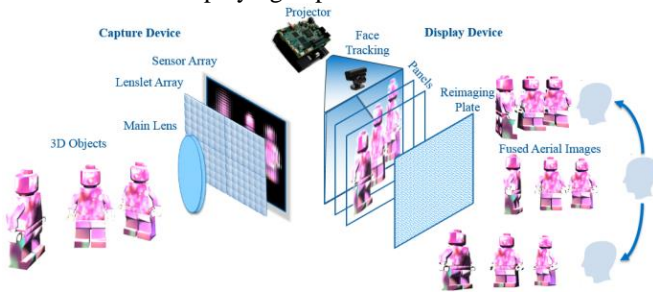

Fig. 1: The light field capturing (left) and multi-panel displaying devices (right).

Multi-panel displays have been implemented with solid-state components to render depth planes in a volumetric display without involving moving parts, which are prone to failure over time. Depthcube [5] utilizes a time multiplexing scheme with a fast projector synchronized with liquid crystal (LC) shutters. In order to support parallax on this display, 20 panels are placed 5 mm apart in a 4 inch depth volume. It is possible to retain depth information and reduce the number of panels to 3 using time-multiplexed projection techniques and a lens array [9] but the resolution is compromised. In other variation, multiple translucent display layers at different depths are combined onto the same optical path where each of the layers is comprised of multiscopic elements (e.g. parallax barrier or lenslet array) which emit true view dependent rays [10]. Fast switching polarization devices have also been used in tandem with projection and polarized films to render 3 depth planes [11]. Recently an inexpensive multi-panel system [12] divided the vertical resolution into pre-distorted strips, which are rendered onto 10 panels 1 cm apart without utilizing time multiplexing.

Stacked LCD panels [13, 14] are typically limited to three panels due to Moiré effects and light attenuation by the polarizers and LCD shutter stacks. These have been utilized to demonstrate depth effects and blending [15] on in-plane-switching and vertical-alignment panels when the polarizers are removed and the LCDs are rotated with respect to each other between 15 and 30 degrees. These displays have been utilized commercially in entertainment and automotive industries [16]. The content rendered on these systems is typically a virtual model designed to fit in the volume of the depth planes, sliced and blended at the edges of the model in order to mask discrete edges of content shared between the planes from the viewer.

For the last 20 years eye tracking systems [17-19] have been paired with auto-stereoscopic displays to optimize the rendered content. We present an example system (see Fig. 8) that combines face tracking with three time-synchronized LC shutters, a projector, and re-imaging mirrors. These components were chosen to maximize the perceived presence of a mid-air rendering while utilizing static components.

Recent developments of LF cameras have made it possible to render real-world LF content on such displays. Our LF retargeting pipeline is introduced to expand the range of content which can be rendered, propose an approach to boost parallax based on estimated depths, and combine this with tracking to render views with a wider viewing zone.

The supported viewing zone and perceived depth of the system depends on the separation of the panels, the amount of blending, the delivered parallax, and the viewing range allowed by the reimaging plate. In practice, panels should not be separated by more than 25 mm on a 480 mm diagonal display when viewed at a comfortable viewing distance because greater separation limits the range of parallax support. In our system, calibration of aerial images from certain points for each panel was required to correlate the line of sight of a viewer with the fused image alignment.

## 3. LIGHT FIELD RETARGETING PIPELINE

The LF retargeting pipeline is a comprehensive set of algorithms to synthesize LFs with enhanced parallax content from the LF capture stage and render them in a compelling, interactive way on a multi-panel display stage. The parameters driving the algorithms (e.g. parallax scale, slicing density, filling procedure, blending type, view-dependent calibration) were chosen by soliciting user feedback on maximizing depth perception, motion parallax, visual clarity, image fusion, and responsiveness. We collected feedback from over 20 people throughout the design process. A block diagram highlighting these algorithms is shown in Fig. 2.

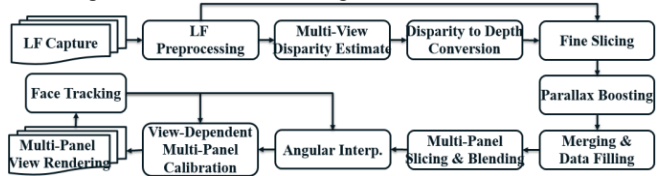

Fig. 2: The light field retargeting pipeline for multi-panel displays.

### 3.1. Light field capture and preprocessing

The retargeting pipeline starts by capturing real-world LF through one of the following three methods: 1) a single-aperture LF camera with limited parallax content such as single-shot plenoptic camera [3, 21], 2) a multi-shot focal stack capture [22], or 3) multi-camera arrays with small [23] or large [24] baselines. While our LF retargeting is more useful in cases where the small aperture limits the amount of captured parallax, it still provides controllable parallax content in post-processing for all captured LFs. Once captured, basic preprocessing operations are performed such as extraction of raw view images in rectangular grid format, denoising, color correction, undistortion, and rectification.

### 3.2. Multi-view disparity estimate

Stereo matching algorithms generally do not work well (if applied without further processing) in case of extremely narrow baseline light field images (e.g., plenoptic cameras), even if the applied algorithm is ranked high in the Middlebury stereo matching benchmark [25] because the one pixel disparity error is already a significant error. Recent published papers [26-28] compute reasonably good sub-pixel disparity but only for the center view and at the high computational cost (e.g. the reported runtime of [26] is 6 minutes for a single view while in [28] it is 10 minutes). Lytro power tools [29] also provides the depth map for one view point. Navarro [30] provides a solution to calculate disparity maps for 9×9 LF views with runtime of 29.28 minutes. Our LF retargeting procedure requires depth estimation for all LF views which cannot be efficiently met by any existing approach.

In order to efficiently calculate accurate subpixel disparity for all LF views (usually more than 100) from a densely sampling LF camera, we propose a framework of multi-view subpixel disparity estimation, shown in Fig. 3. Our framework can for instance generate disparity maps for all 14×14 LF views (24 reference views + 172 propagated ones) captured by Lytro camera in about 11 sec (running on an Intel i7 3GHz 8-cores machine).

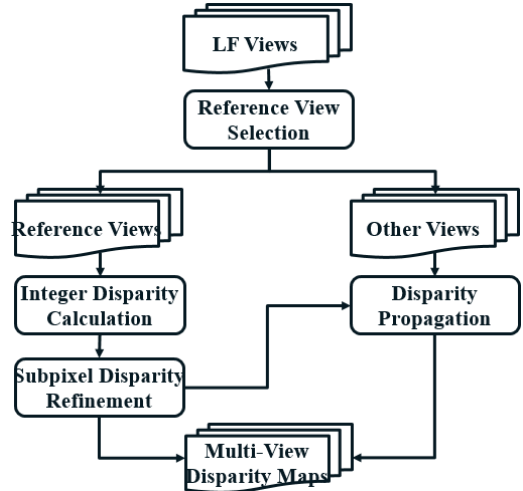

Fig. 3: Framework of multi-view subpixel disparity estimate.

This estimate method functions as follows; given all LF views corresponding to varying perspectives of the aperture sampling, select a subset of views as references for disparity calculation. The principle of view selection is to subsample views on both horizontal and vertical directions to avoid occlusion. The number of skipped views is a tuned parameter, to balance the computational cost and at the same time avoid big occlusion. Fig. 4 on the left shows an example of 16 selected reference views from 14×14 LF views.

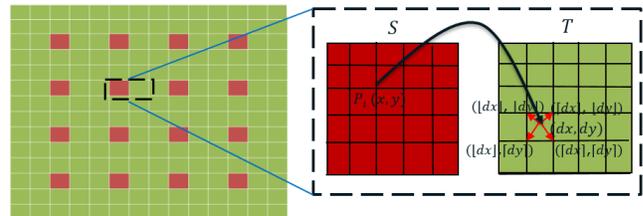

Fig. 4: Example of selected reference views shown in red color (left) and illustration on how to propagate their values to the neighboring views (right).

For a selected reference view, we choose a subset of cross hair views on both horizontal and vertical coordinates for disparity calculation such as shown in left-top corner of Fig 5. Cross hair camera view selection takes advantage of the multiple views given by the light-field which can handle occlusion (i.e., a common problem with stereo pairs), reduce noisy match and at the same time avoid expensive computational cost (i.e., if all 14 by 14 views used).

Once the set of view pairs is selected, a multi-baseline disparity algorithm estimates disparity (i.e. for reference views) using a two-step approach. First, the integer disparity is calculated at a coarse level using a light-weight process, and then is refined to subpixel precision within a small range. Our multi-baseline integer disparity calculation is based on stereo disparity algorithm [31]. The estimating algorithm flow shown in Fig. 5 can be described as follows:

- First, to have more robust pixel correspondence match, we extract features such as gradient, census etc. to represent a pixel.
- An adaptive shape support region instead of a fixed size region is desired for accurate disparity estimates, therefore only the pixels of the same depth are used for sum of absolute difference (SAD) calculation. Fixed SAD window size failed at low texture neighborhood and depth discontinuity areas. To find the adaptive shape support region, each pixel (x, y) will extend to four directions (left, right, up and down) until it hits a pixel that the color, gradient or grayscale difference between this pixel and pixel (x, y) is beyond certain thresholds.
- For each candidate disparity d ∈ [0, maxD] and for each image pair, we initialize the cost per pixel using the absolute difference (AD) between pixel's feature. For each image pair, aggregate the AD errors of all pixels in the support region S using a sum of absolute difference (SAD). This aggregation can be efficiently calculated using integral image techniques.
- We then aggregate cost cross all image pairs using SUBMIN method, which is the minimum cost summation from any subsets of 3 image pairs out of the four pairs in Fig. 5. SUBMIN has the advantage over just taking the minimum cost from four pairs, which is likely to be biased by matching noise from any single pair. SUBMIN also has the advantage over summing costs from all four pairs, which tends to be impacted by occlusion errors from any single pair.
- Finally, we will check the variance of the cost curve epi(d), d ∈ [0, maxD]. If the variance is too low, it means there is no unique correspondence for that pixel. If the cost curve passes uniqueness check, we use the winner-take-all (WTA) approach. A given pixel (x, y)'s integer disparity in the reference camera is chosen by finding the minimum d in the cost curve; epi(d), d ∈ [0, maxD].

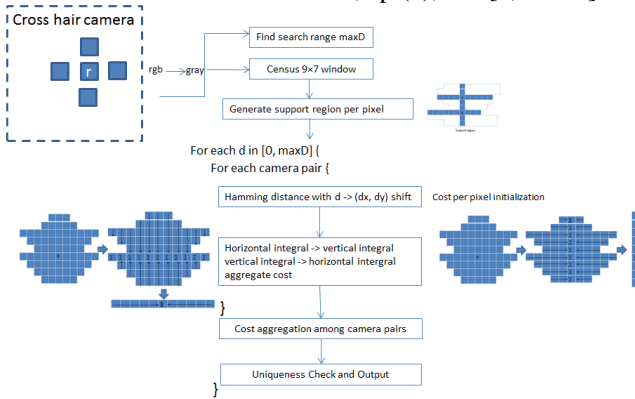

Fig. 5: Multi-baseline integer disparity estimation.

With estimated integer disparity, our multi-baseline subpixel disparity refinement algorithm refines the output in a small neighborhood of the integer disparity as illustrated in Fig. 6. Images and integer disparity are up-scaled by 2X. Support region is also up-scaled by 2X. The feature extraction from the up-sampled images is done using the same method in the integer disparity calculation. For each pixel Pi and its integer disparity di, the disparity search range is around the integer disparity: [di-μ, di+μ]. For each candidate disparity d ∈ [di-μ, di+μ], we compute individual pixel's cost, aggregate all pixels' cost in its support region, and aggregate cost cross all image pairs similar to the procedure in integer disparity calculation. At the end, we get a cost curve of {epi(d), d ∈ [di-μ, di+μ]}.

We used a similar idea of [32] to interpolate cost curve epi(d) to generate denser samples - ēpi(d). In order to find the minimum d' in the interpolated cost curve, we first calculate d'= $\arg\min_d$ ēpi(d), then getting mathematical minimum đ of the curve (the vertex of the curve) by fitting with a parabola y = ax² + bx + c using d' and its two closest points.

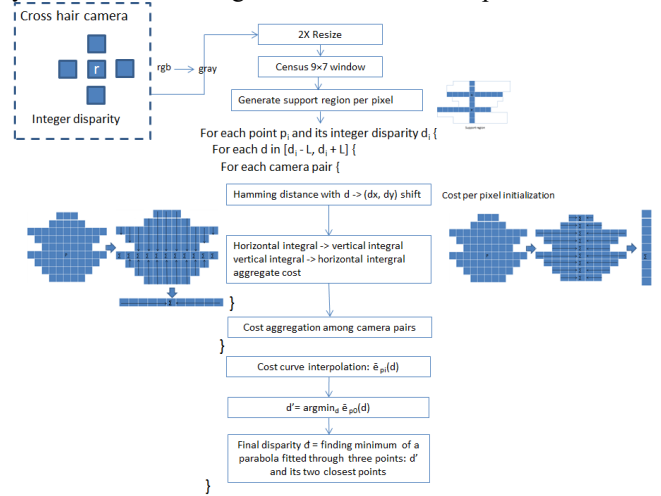

Fig. 6: Multi-baseline sub-pxiel disparity refinement.

Since the search range is in a small range instead of whole search space, we can afford to have an accurate subpixel disparity algorithm to generate disparity in a continuous depth space with 1/20 subpixel accuracy. This two-step approach can significantly reduce runtime without sacrificing precision compared to running an expensive sub-pixel disparity algorithm directly.

The baseline between LF views from a single-aperture LF camera is small, therefore pixels between LF views share high redundancy. We can use this property to propagate disparity calculated from reference views to other views using weighted forward remapping. Assume there are a set of reference views {S} whose disparities have already been calculated, we will propagate their disparities to an unknown view T. For each reference view S, a pixel $P_i$ (x, y) maps to a pixel (dx, dy) in target view T based on $S_D(x,y)$ - disparity value of pixel (x,y) of S and relative view position between S and T, see Fig. 4 - right.

Since $S_D(x,y)$ is a subpixel disparity value, mapped pixel (dx, dy) is very likely on non-integral coordinates, as shown in right image of Fig. 4. The disparity value of pixel $P_i$ (x, y) in S contributes to four integer pixels(⌊dx⌋, ⌊dy⌋), (⌈dx⌉, ⌊dy⌋), (⌊dx⌋,⌈dy⌉) and (⌈dx⌉,⌈dy⌉)near (dx, dy) in T, and weighted by: a) distance from (dx, dy) to these four integer pixels. For example, the upper-left pixel(⌊dx⌋,⌊dy⌋)'s distance to pixel (dx, dy) is alpha = dx – ⌊dx⌋  horizontally, and beta = dy – ⌊dy⌋  vertically. Its distance weight = (1.0-alpha)×(1.0-beta). The closer it is, the higher the weight is; b) color similarity of source pixel (x, y) in $S_i$ to four integral pixels in T. There could be multiple reference images {S} mapped to a target disparity T. Some pixels in T might be occluded in certain images of {S} but visible in others. By considering the color similarity, we can have a more robust remapping to deal with noise and pixel occlusion. The higher the color similarity is, the higher the weight is. The total weight of $P_i$ in S contributes to the target pixels in T is the multiplication of distance weight and color weight.

### 3.3. Disparity to depth conversion

Given an estimated disparity map d, the depth map z can be calculated by $z = \frac{bf}{d+d_o}$, where b is the baseline and f is the focal length of the LF camera. $d_0$ refers to the zero disparity plane which may not be associated with an infinite depth since LF cameras can be configured to focus at any depth planes. We can calculate $d_0$ and   bf using the following equations:

$$d_0 = \frac{\min(z)\max(d) - \max(z)\min(d)}{\max(z) - \min(z)} \quad (1)$$
$$bf = \max(z)(\min(d) + d_0) \quad (2)$$

where min(z) and max(z) are corresponding to the minimum and maximum distance in the captured scene. These are camera and scene dependent. Once z is calculated, we then normalize it between {0,1} where 0 refers to the closest distance and 1 is the farthest distance (background).

### 3.4. Fine slicing

Afterward, the estimated multi-view depth maps are uniformly quantized into a pre-set number of levels and the LF views are sliced accordingly such that pixels close together in depths (i.e. belonging to the same quantization level) are brought together on same slice. We have adopted a simple slicing approach in favor of an efficient lightweight retargeting. Having more slices will result in a better parallax boosting and data filling however this is upper bounded by the subpixel accuracy of the estimated depth maps and requires more computation.

### 3.5. Parallax boosting

The fine slices for all views are then translated in proper directions and magnitudes relative to their normalized angular and depth values. This is done with respect to a reference view $Ref_{V_{x,y}}$ and a reference depth plane $Ref_D$ according to the following equations:

$$T_{x_{i,k}} = \lfloor (Ang_{x_i} - Ref_{V_x})(QuantD_k - Ref_D)Scale \rfloor \quad (3)$$

$$T_{y_{j,k}} = \lfloor (Ang_{y_j} - Ref_{V_y})(QuantD_k - Ref_D)Scale \rfloor \quad (4)$$

Where $Ang_x$ and $Ang_y$ are the normalized angular coordinates in the range {-0.5, 0.5} indexed by i and j = 1, …, number of views in one dimension, QuantD is the normalized quantized depth map with values between {0,1} (where 0 refers to the closest object) indexed by k = 1, …, number of slices, Scale is the parallax boosting amount, and ⌊ ⌋ brackets are rounding to the nearest integer for efficient filling results.

Note that according to these equations, the slices of the reference view or those located at the reference depth plane will experience no shifts while those at extremist view from the reference view and farthest distance from the reference depth plane will experience the largest shifts.

The user can control the max shift Scale in pixels, which is upper bounded by the physical viewing zone of the display. Adjusting this parameter results in stretching and compressing effects in the depth of field perceived on the display. Note that Scale = 0 implies that no parallax boosting or filling is done and the LF content is displayed in the same manner it was captured. Additionally, the user can select a reference depth plane $Ref_D$ to whom all slices will be shifted with respect to. This mimics having the eyes fixating on certain depth plane. For instance, $Ref_D = 1$ means the eyes are fixating on the background hence slices are shifted with respect to background, which remains fixed. Similarly, $Ref_D = 0.5$ corresponds to shift with respect to center while $Ref_D = 0$ is used for shift with respect to front layer.

### 3.6. Merging and data filling

After translation, the slices per view are merged together such that the upper layers may overwrite the back ones to support occlusion as seen from the observer's position. The parallax boosting and merging operations introduce new occlusion relations making the synthesized views appear as if they were captured by virtual cameras of larger baseline. However, these new occlusions result in holes (black regions) that require data filling.

We constrained the slices' shifts to integer values for computational efficiency so that intensity values at sliced boundaries may not be spread over neighboring pixels. This is critical to get good light-weight filling results in the black regions since the boundary pixels will be utilized to interpolate these missing data. Note that the rounding to integer shifts in parallax boosting's equations do not compromise the subpixel accuracy of the estimated disparity map since it is being applied on the boosted values and not on the depth map directly. A nearest interpolation is used for the data filling so the sharpness is maintained in the filled regions. The filling process is then followed by median filtering within a small window (3×3) to provide consistency in the filled regions.

Similar to the processed LF, the estimated multi-view depth maps undergo same fine slicing, translations, merging,

and filling to keep tracking the depth values of the views synthesized with enhanced parallax. This is needed for the multi-panel slicing and blending stage.

### 3.7. Multi-panel slicing and blending

Despite the panels are separated physically by a fixed distance, we choose to slice the LF content in a non-uniform fashion to efficiently fit the content on the display. For instance, objects may not be populating the captured scene in depth homogeneously, hence a uniform multi-panel slicing may result in panels without any content leading to a waste in the depth budget. The non-uniform slicing is executed based on the synthesized depth maps using a standard multi-level Otsu thresholding algorithm [33]. Other thresholding techniques can be utilized as well such as K-means clustering [34] or histogram-based equal-counts thresholding (i.e. finding thresholds in the histogram that result in slices of equal-pixel counts) but we selected Otsu thresholding since it delivered slight better results for our displayed data.

After optimizing the quantization levels (representing panels), the content has to be blended across the panels to impose continuity in depth despite the coarse sampling of the panels. All-panel linear blending is applied so the intensity of each pixel per synthesized view is shared across all panels weighted by the normalized distance to these panels. Despite the non-uniform depth thresholding assigned to the panels, the correct depth perception is preserved by the all-panel blending step. All-panel blending is further illustrated in Sec. 4.3.

### 3.8. Face tracking and angular interpolation

We utilize an open source face-tracking algorithm [35] that detects and tracks certain facial features with a simple camera and returns the angular information (e.g. viewer's position) with respect to the camera's center.

Since the captured LF views are discrete in the angular domain while the returned tracking measurements are continuous, novel blended views have to be synthesized from the neighboring ones to accurately render the proper view correspondent to the viewer's position. We use bilinear interpolation in the angular domain for this purpose. This helps making seamless transitions of the views on the display as the viewer is changing position.

### 3.9. View-dependent multi-panel calibration

In the display prototype we built, the projector's axis is not aligned with the center of the panels besides the projected beam gets diverged during its propagation through the panels. This means that the content projected onto the panels may no longer be aligned and the amount of misalignment may differ according to the viewer position.

To maintain alignment, the display has to be calibrated first. One of the panels with certain view (in our setup, the back panel with the central view) is selected to be a reference whom the content of all other panels for all views are aligned to. Then given a calibration pattern, we empirically adjust scaling ($S_x$, $S_y$) and translation ($T_x$, $T_y$) parameters then apply this transformation on the content of other panels (one at a time) for a single viewing position till their content gets align with the pattern on the reference panel. Note that the scaling parameters implicitly spatially upsample the captured LFs to fit the spatial resolution of the projector utilized in the multi-panel display. This calibration process is reiterated for selected viewing angles (covering field of view ~ 19º × 19º which is 30cm×30cm at 90 cm distance in our setup, see Fig. 7) to find calibration parameters per panel per view. These calibration parameters are then passed into a linear fitting polynomial to learn its coefficients and use it later to derive the calibration parameters at any viewing angle.

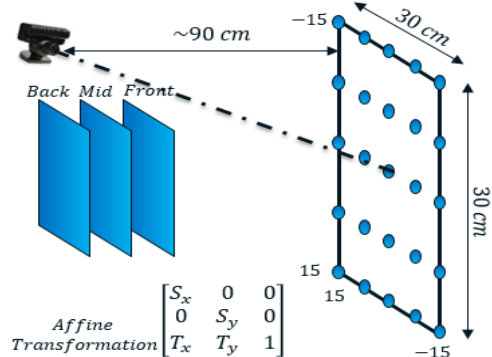

Fig. 7: Calibration geometry in three-panel display for coarse sample of points representing viewer's position and camera oriented toward the central point.

For a given viewer's position, the interpolated view undergoes a set of affine transformations (one for each panel) with calibration parameters derived from the fitted polynomial. This is done on-the-fly interactively with the viewer's position to impose alignment (in scale and translation) on the rendered content for all panels. In case the relative position of the projector and the panels changed (due to mechanical movement), the entire calibration procedure has to be repeated which takes about 45 min for 5×5 preselected calibration points. A comparison of retargeted content as seen on the display with and without calibration is given in Sec. 4.4.

### 3.10. Multi-panel rendering

The calibrated slices for the sensed view are then projected onto the correct panels synchronized with the projector at refresh rate sufficient to perceive 3D content without flickering. This is achieved by synchronizing the VSYNC output from the projector with rear projected liquid crystal panels, which can switch in 750 microseconds between a transparent and clear state.

Using an Arduino Pro microcontroller, the rising edge of the VSYNC output of the lightcrafter 4500 projector from TI

in pattern sequence mode with a pattern exposure of 5270 microseconds and a pattern period of 5555 microseconds. The output located on pin T2 is used to trigger an interrupt linked to a function that initiates a transition between shutters which drops the voltage for the next panel and 400ms later raises the voltage for the current panel. Like two curtains transitioning on a performance stage, projection light cannot pass through both panels in the clear state or it will be perceived as a hotspot. The color content is upper bounded by the projector's speed and the panels' response time. In our setup which uses three panels, we had to operate in a single color mode (white) in order to deliver a refresh rate for 3 panels at 180 Hz.

Finally, we loop back to the face tracking stage for the next measurements and update the content interactively on the multi-panel display.

## 4. RESULTS AND DISCUSSIONS

In this section, we demonstrate the impact of various stages in the LF retargeting pipeline on various LF datasets. Unless stated differently, the retargeting parameters used in the study were set as follows: 24 reference views for the disparity estimate, 100 fine slices, $Ref_V = (0,0)$, eyes' fixation on the central panel with $Ref_D = 0.5$, parallax scale = 100, all-panel blending, and multi-panel calibration. The LF data shown in the first four sections was captured by Lytro Illum [36], an example of commercial plenoptic camera, at 14×14×375×541 angular-spatial resolution. The last section demonstrates the retargeting results for LFs generated from different resources. The input LF data is then retargeted for three-panel display embodying a DLP D4500 projector of resolution 1140×912 running at 180 Hz with 7-bit grayscale pattern rate. All final retargeting results where verified visually in the actual setup, shown in Fig. 8. However to facilitate the analysis we present the results after each stage on selected corner views where retargeting effects are maximized. In terms of timing (for Lytro-captured LFs), multi-view disparity estimate takes about 11 sec, the stages from fine slicing till all-panel blending are executed in ~250 ms, and the interactive loop including the face tracking, the angular interpolation, and the multi-panel calibration is executed at 15 ms on Intel i7 3GHz 8-cores machine.

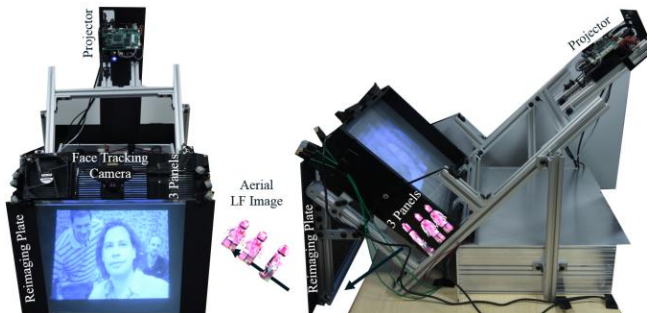

Fig. 8: Experimental setup for multi-panel display showing front and side views along with actual retargeted central LF view as seen on the display.

### 4.1. Efficiency of the multi-view disparity estimate

In this study, we evaluate the efficiency of the multi-view subpixel disparity estimate in terms of quality and runtime as we are tuning the number of reference views considered during the estimate. For 14×14 LF views, we consider 13, 24, and 37 reference views and present the estimated disparity maps of two corner views in Fig. 9. The results match our intuition that with higher number of reference views for disparity calculation, we can recover higher depth resolutions (e.g. recover multiple depth layers on facial profile and human body). However, more reference views result in more computations and longer runtime. Note that for 24 reference views, we achieve a good balance of depth resolution and runtime.

| | top-left view | bottom-right view | time |
|---|---|---|---|
| 13 | | | 7.53 sec. |
| 24 | | | 11.01 sec. |
| 37 | | | 18.27 sec. |

Fig. 9: Evaluating the efficiency of the multi-view disparity estimate as a function of number of reference views (across rows) at two corner views (across columns). The runtime to estimate disparity maps for all views is shown in last column in seconds.

### 4.2. Retargeting with and without boosted parallax

The limited parallax content in LFs at the capture stage results in shallow depth content and high angular redundancy since views projected on the multi-panel display exhibit slight changes while viewer is moving left to right or top to bottom within the display's viewing zone. At the boosting parallax stage, the views are synthesized to provide more angular content by increasing the virtual baseline. The amount of change is controlled by $Step$ parameter to compress or stretch the perceived depth on display. Figure 10 illustrates the differences in two corner views when boosting parallax or not. The correspondent disparity maps are provided as well for a better visualization. Note that there is barely any change in case of original data with no parallax as opposed to those with enhanced parallax (e.g. note the difference in occlusion relations between the two persons at the front and the back of the scene).

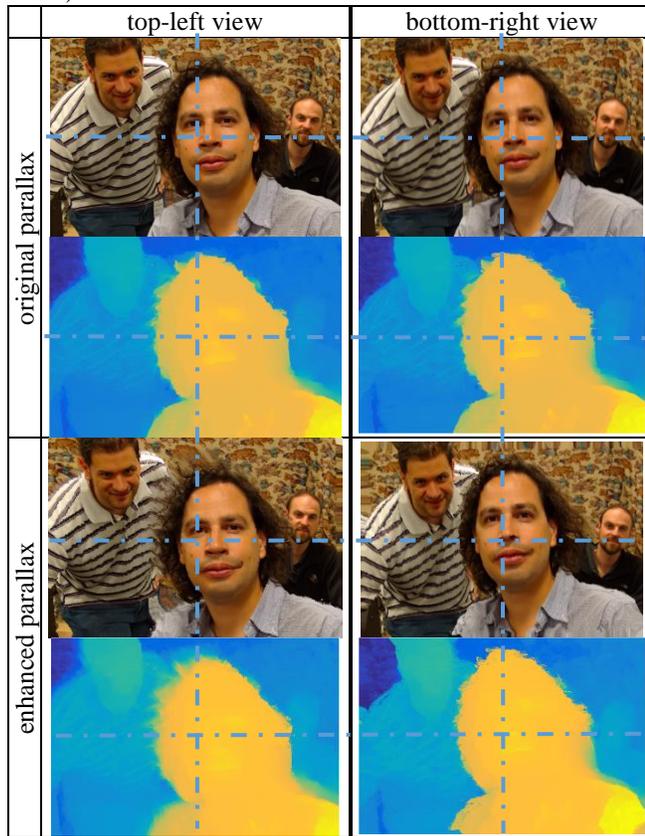

Fig. 10: Visual comparison of LF retargeting and the correspondence disparity maps with and without boosted parallax (across rows) at two LF corner views (across columns). The horizontal and vertical lines overlaid on the views aim to help illustrating the differences.

### 4.3. Retargeting with and without blending

Blending techniques enable an efficient implementation of multi-panel displays by relaxing the required number of panels while maintaining continuity in depth. In this study, we shall compare three cases; no-blending, two-panel blending and all-panel blending implemented for three-panel display. In the no-blending case, the intensity per pixel is mapped totally to the closest quantized panel. In the two-panel blending case, the intensity of each pixel is shared between the sandwiching quantized panels with normalized weights set by distance to these panels. Note that pixels with depths before the first quantized panel or after the last quantized panel are mapped to single panel. In all-panel blending case, no matter where the pixel is located its content will be weighted and shared on all-panels. Figure 11 illustrates the difference between these cases for two pixels. Note that the coarse depth slicing is implemented non-uniformly to efficiently utilize the panels.

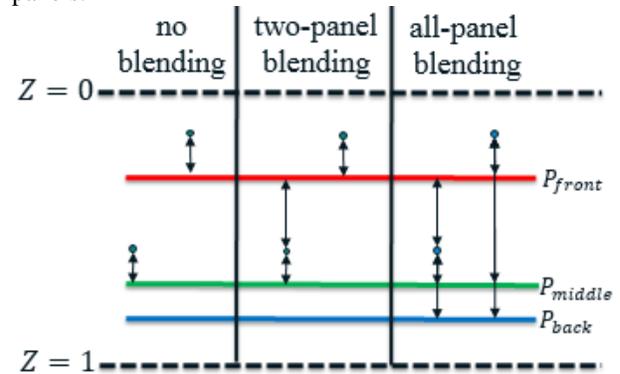

Fig. 11: Sketch illustrating various blending cases for two pixels where one of them is located in front of the front panel and the other is sandwiched between the front and middle panel in three-panel display. The arrows indicate the panels sharing the intensity of a given pixel.

A visual comparison of the blending cases is presented in Fig. 12. It is clear from Fig. 12 - 1$^{st}$ row how the no-blending case can result in discretized layered appearance in depth. On other hand in the two-panel blending case in Fig.12 -2$^{nd}$ row, we can see that pixels on both middle and back panels are well blended; however, those on front panel are not. Finally, in all-panel blending case seen in Fig. 12 -3$^{rd}$ row, the intensity of pixels are distributed on all panels, resulting in well-blended views.

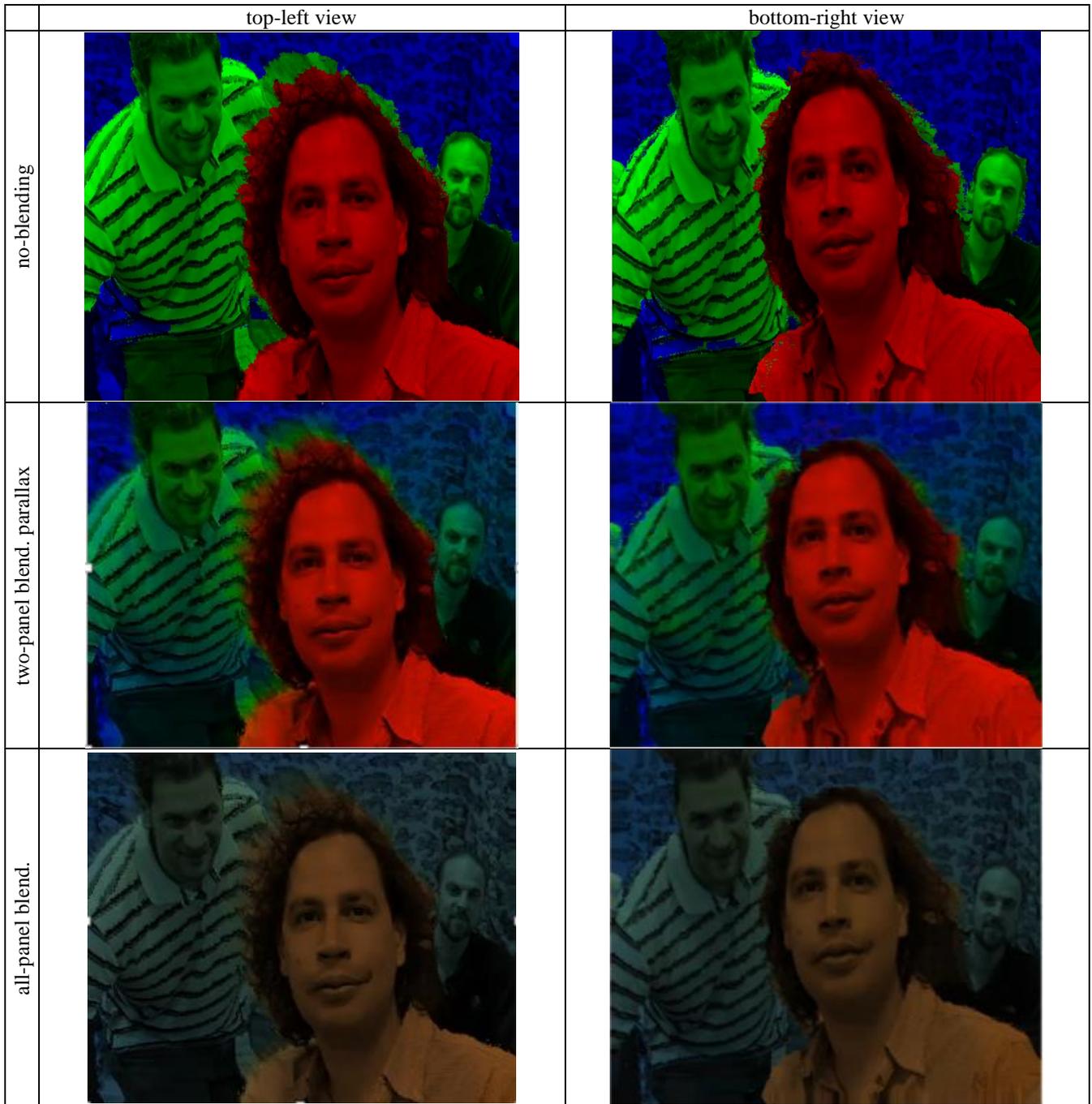

Fig. 12: Visual comparison of LF retargeting with various blending types (across rows) at two LF corner views for three-panel display. Note that the color channels are used to represent the content of the three panels with red representing the front panel, green for middle panel, and blue for back panel.

### 4.4. Retargeting with and without calibration

Calibration is required to preserve alignment of content on all panels by correcting for scale and translation given a pre-calibrated linear fitting polynomial. In Fig. 13, we share the final LF retargeting results as inputted on projector and seen on the display with and without calibration for two different corner views. Note that after applying calibration, the panels' content of projector's images are not any more aligned (Fig. 13 – 3rd row) as opposed to those before the calibration (Fig. 13 – 1st row). A comparison in Fig. 13 between the 2nd and 4th rows (both captured by DSLR camera set at the exact position for a given view) reveal how critical the multi-panel calibration is to fuse the content across all panels for a given view.

| | | top-left view | bottom-right view |
|---|---|---|---|
| without calibration | on projector | 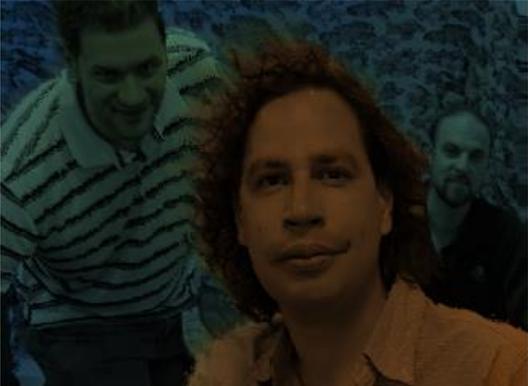 | 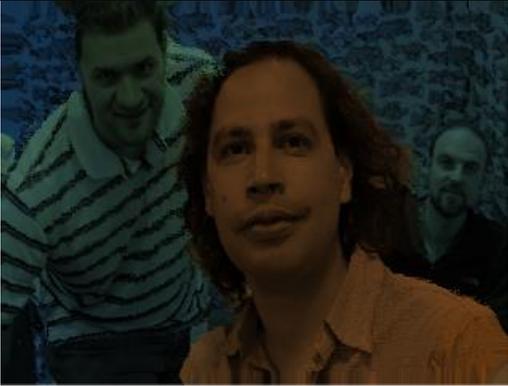 |
| | on display | 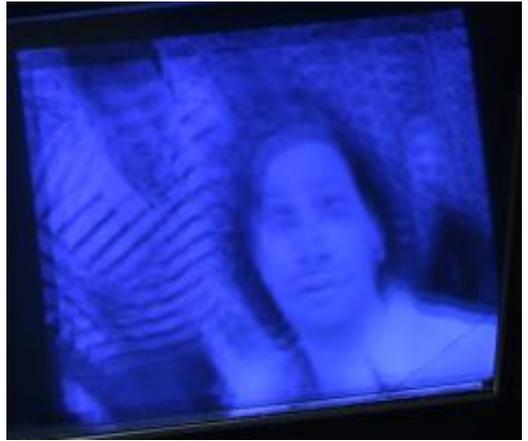 | 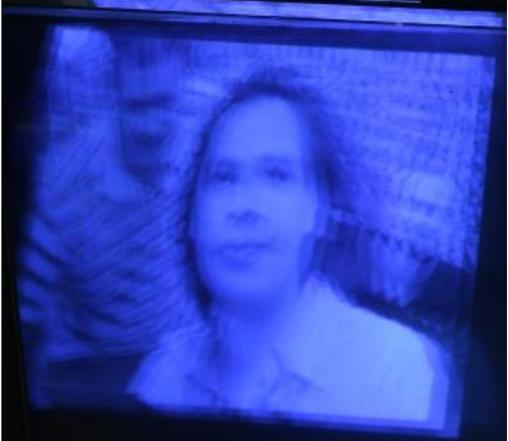 |
| with calibration | on projector | 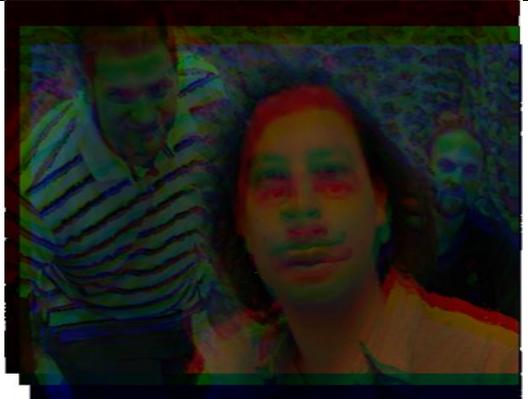 | 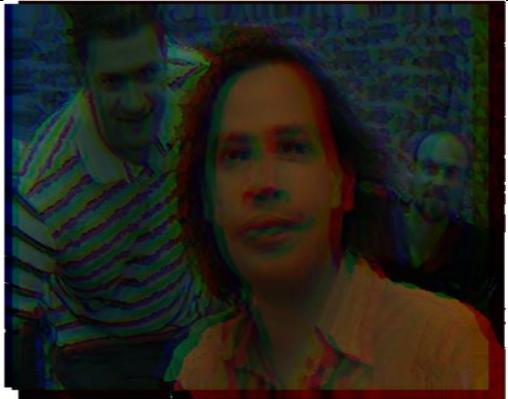 |
| | on display | 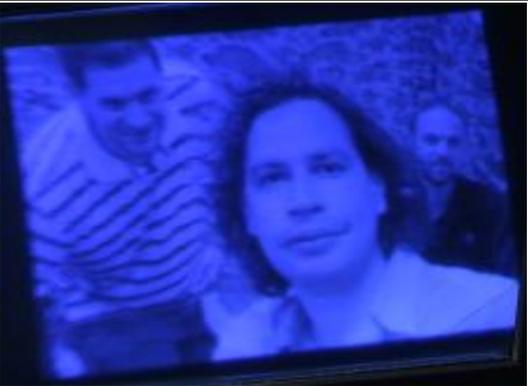 | 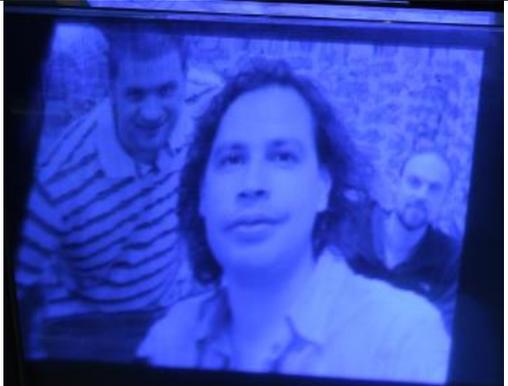 |

Fig. 13: LF retargeting results as inputted on projector and seen on display with and without calibration (across rows) at two different corner views (across columns).

### 4.5. Retargeting LFs from various input resources

To further validate the retargeting pipeline for various input LF resources, we consider four additional LF samples; the first is a group of people captured by Lytro Illum (Plenoptic 1.0) camera [36] (Fig. 14), the second is an electronic board captured by Raytrix R29 (Plenoptic 2.0) camera [37] (Fig. 15), the rest are synthesized LFs for Papillon and Mona scenes from Heidleberg database [38] (Figs. 16-17). In all cases, we show the two corner views for the original LFs, for the enhanced-parallax LFs, for the calibration results, and for the final retargeting results as seen on the multi-panel display.

## 5. CONCLUSION

The light field retargeting pipeline presented here is a set of generalized algorithms that process light fields captured with limited parallax and realize them in an interactive manner on multi-panel display despite the architectural and dimensional differences between the capturing systems. The retargeted light field views in our example system were guided by viewers in our lab who we asked to describe their experience of the depth, fusion, clarity, and responsiveness of the aerial image. We have identified key considerations for retargeting and present a more efficient and responsive experience enabled by: 1) efficiently estimating multi-view depth maps, 2) parallax boosting and data filling to synthesize views of larger baseline, 3) multi-panel slicing and blending to preserve continuity in depth, 4) on-the-fly angular interpolation to properly render the correspondent view interactively based on face tracking measurements, and 5) multi-panel calibration to fuse the synthesized view content on the panels.

In future work we shall further optimize the light field retargeting algorithms and improve the execution time. We also plan to augment a gaze tracking system to provide feedback in the parallax boosting stage so that the fixation of the eyes factored real-time when synthesizing the novel views. Finally, we will extend our light field retargeting solution to support other inputs provided by RGB-D cameras and 3D virtual objects.

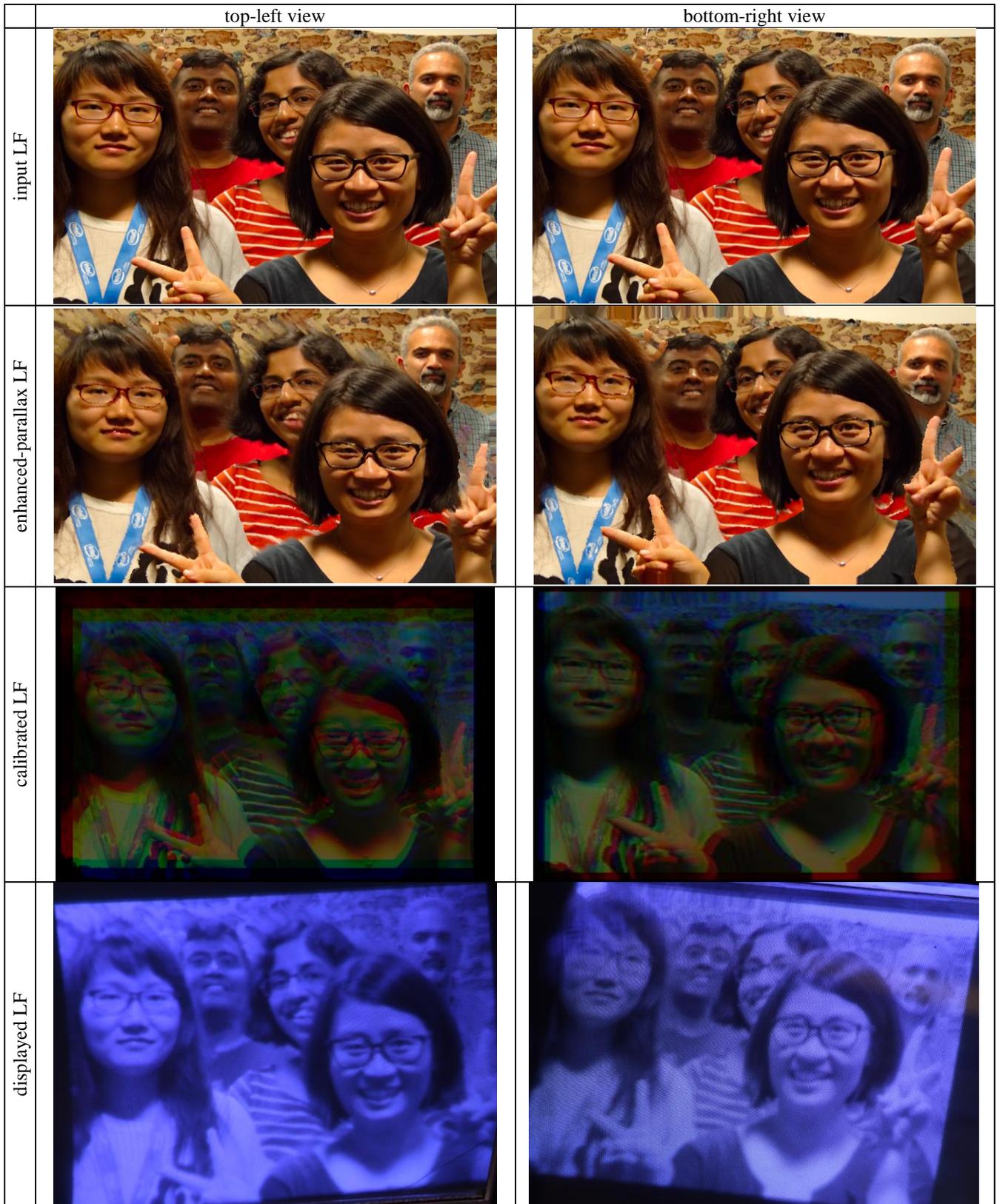

Fig. 14: Retargeting LF captured by Lytro camera at various stages for two corner views.

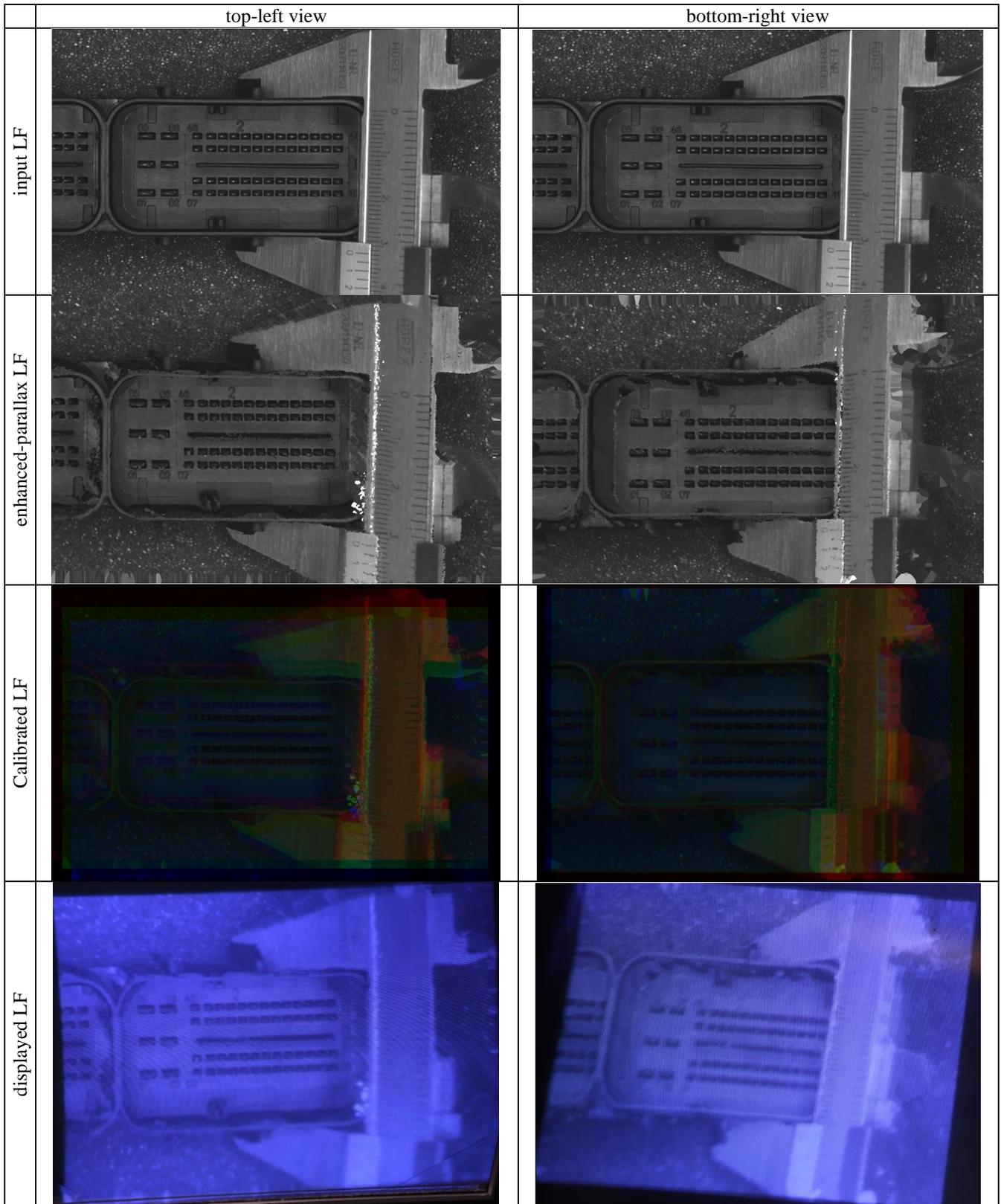

Fig. 15: Retargeting LF captured by Raytrix R12 camera at various stages for two corner views.

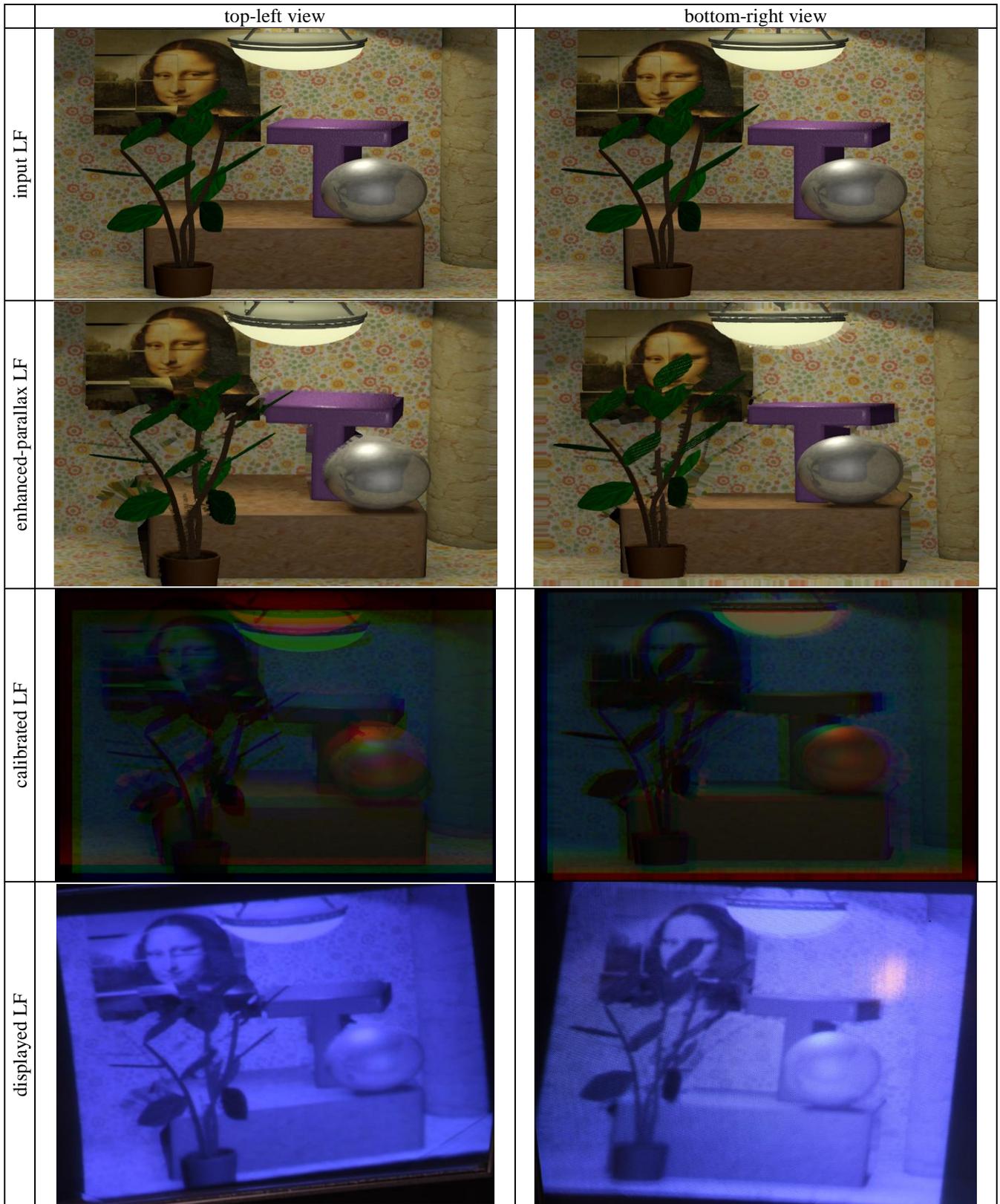

Fig. 16: Retargeting the synthesized Mona LF from Heidelberg database at various stages for two corner views.

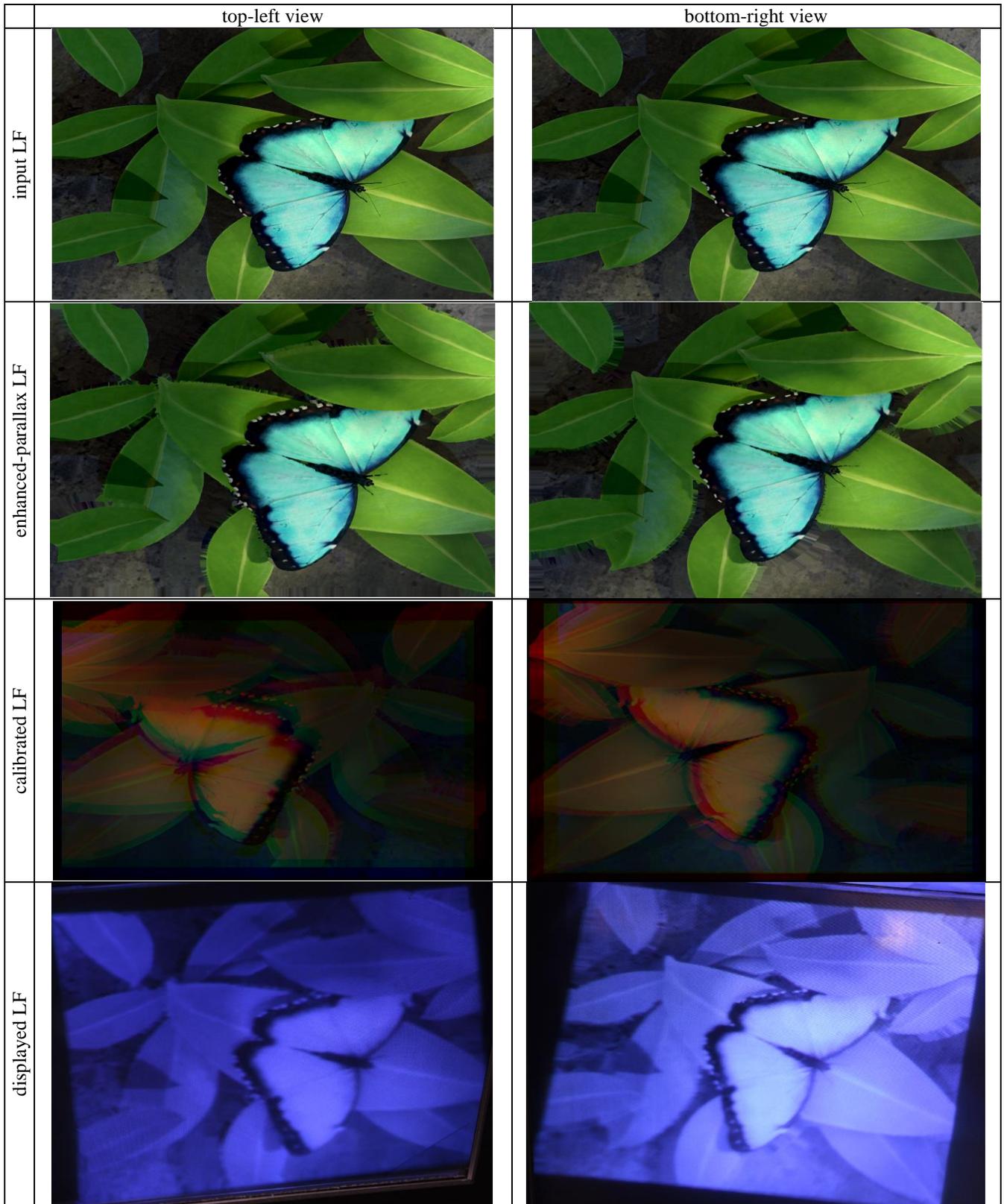

Fig. 17: Retargeting the synthesized Papillon LF from Heidelberg database at various stages for two corner views.